\documentclass[twocolumn,amsmath,amssymb,floatfix,superscriptaddress,prb,longbibliography]{revtex4-2}
\usepackage{lineno,hyperref}
\usepackage{bm}
\usepackage{amsmath}
\usepackage{gensymb}
\usepackage{epstopdf}
\usepackage{booktabs}
\usepackage{longtable}
\usepackage{tabularx}
\usepackage{multirow}
\usepackage{setspace}
\usepackage{mathrsfs}

\usepackage{graphicx}
\usepackage{ulem}
\usepackage{xcolor}
\usepackage{hyperref}
\usepackage{amsfonts} % for "\mathbb" macro

\newcommand{\Z}{\mathbb{Z}}

\begin{document}

\preprint{APS/123-QED}

\title{Al$_2$MnCu: A magnetically ordered member of the Heusler alloy family despite having a valence electron count of 24}% Force line breaks with \\

\author{Soumya Bhowmik}
\affiliation{Condensed Matter Physics Division, Saha Institute of Nuclear Physics,
              A CI of Homi Bhabha National Institute, 1/AF Bidhannagar, Kolkata 700064, India}
\author{Santanu Pakhira}
\affiliation{Department of Physics, Maulana Azad National Institute of Technology, Bhopal, M.P. - 462003, India}
\author{Renu Choudhary}
\affiliation{Ames National Laboratory, Iowa State University, Ames, Iowa 50011, USA}
\author{Ravi Kumar}
\affiliation{Atomic \& Molecular Physics Division, Bhabha Atomic Research Centre, Mumbai 400 094, India}
\author{Rajashri Urkude}
\affiliation{Beamline Development \& Application Section, Bhabha Atomic Research Center, Trombay, Mumbai 400085, India}

\author{Biplab Ghosh}
\affiliation{Beamline Development \& Application Section, Bhabha Atomic Research Center, Trombay, Mumbai 400085, India}

\author{D. Bhattacharyya}
\affiliation{Atomic \& Molecular Physics Division, Bhabha Atomic Research Centre, Mumbai 400 094, India}

\author{Maxim Avdeev}
\affiliation{Australian Nuclear Science and Technology Organisation, Locked Bag 2001, Kirrawee DC, New South Wales 2232, Australia}
\affiliation{School of Chemistry, The University of Sydney, Sydney, New South Wales 2006, Australia}
\author{Chandan Mazumdar}
\email{chandan.mazumdar@saha.ac.in}
\affiliation{Condensed Matter Physics Division, Saha Institute of Nuclear Physics,
              A CI of Homi Bhabha National Institute, 1/AF Bidhannagar, Kolkata 700064, India}

\date{\today}

\begin{abstract}

The magnetic property of the Heusler alloys can be predicted by the famous Slater-Pauling (S-P) rule, which states the total magnetic moment ($m_t$) of such materials can be expressed as $m _t\,=\,(N_V-24)\,\mu_B/f.u.$, where $N_V$ is the total valence electron count (VEC). Consequently, no Heusler alloys having VEC = 24 are theoretically expected as well as experimentally reported to have any magnetic ordering. Recently, a special class of Heusler alloys with 50\% concentration of $p$-block elements (anti-Heusler) have been identified, although none of such reported compounds belong to the VEC 24 category.
 Here, we report a new anti-Heusler alloy,  Al$_2$MnCu, that undergoes long-range ferromagnetic (FM) ordering with $T_{\rm C}\sim$315~K and a large magnetic moment of $\sim$1.8 $\mu_B$/f.u. despite having VEC 24. A phenomenological model based on  molecular orbital hybridization is also proposed to understand the magnetism and unusual deviation from the standard S-P rule.
\end{abstract}

\maketitle

\section{Introduction}
Heusler alloys are one of the promising classes of materials to explore for their innumerable technological prowess encompassing a vast range of state-of-the-art characteristics, $viz.$, half-metallic ferromagnet~\cite{PhysRevLett.50.2024,Jourdan2014,PhysRevB.106.115148,PhysRevB.108.045137},
\begin{figure}[]
\centering
\includegraphics[width=0.45\textwidth]{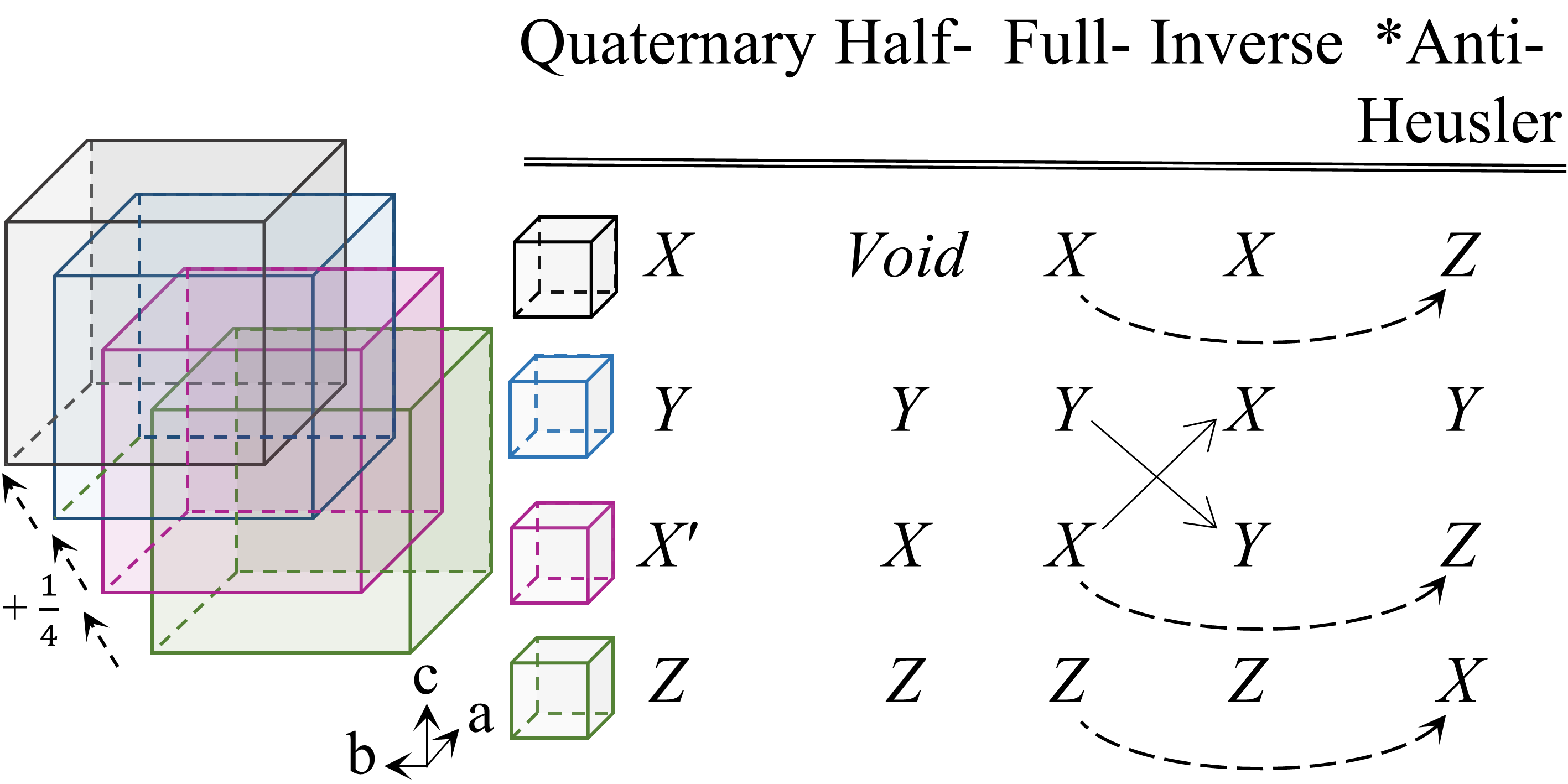}
\caption{Crystal structures of different classes of Heusler alloys having perfectly ordered atomic arrangements. Inverse Heuslers are formed by position-swapping of $X$ and $Y$ atoms (solid arrows) in full-Heusler system, while dashed-arrows indicate the atom replacement scheme (between $Z$ and $X$) to form anti-Heuslers.}
\label{fig:Heusler alloy Crystal structure}
\end{figure}
spin-gapless semiconductor~\cite{PhysRevLett.110.100401,PhysRevB.91.104408,PhysRevB.97.054407}, magnetic skyrmion~\cite{Nayak2017}, topological property, Weyl semi-metal~\cite{Manna2018,doi:10.1073/pnas.1810842115}, spin glass~\cite{PhysRevB.108.054405,PhysRevB.107.184408}, giant magneto-caloric effect~\cite{Liu2012}, thermoelectricity~\cite{PhysRevB.98.205130,10.1063/5.0043552,adfm.201705845}, giant exchange bias~\cite{Nayak2015,PhysRevLett.110.127204}, superconductivity~\cite{sciadv.1500242} $etc.$ Besides the technological perspective, in the realm of basic sciences too Heusler alloys offer an expansive opportunity to study different novel physical properties arising out of the complex interplay of structural and electronic degrees of freedom, as very few classes of materials are known to manifest such versatility. The chemical diversity and the crystal structure in these materials allow for exploring newer compounds with tailored properties through elemental and structural tunability~\cite{PhysRevB.72.184434,PhysRevLett.102.246601,GRAF20111}. The crystal structure of these alloys can be described as a combination of four interpenetrating sublattices, separated by $1/4^{th}$ translation along the diagonal (Fig.~\ref{fig:Heusler alloy Crystal structure}). Each of the sublattices comprises a particular type of atom, $viz.$, $Z$, $X^{\prime}$, $Y$, and $X$, forming a quaternary Heusler ($XX^{\prime}YZ$), where $Z$ is the $p$-block element and rest are transition metals~\cite{GRAF20111,10.1063/1.4959093,PhysRevB.28.1745,PhysRevB.83.184428,PhysRevB.84.144413}. The full-Heuslers ($X_2YZ$) are formed when two of the sublattices (grey and pink) are occupied by the same element ($X^{\prime}=X$), having a higher atomic number than $Y$ ($\Z_X>\Z_Y$)~\cite{heusler1903magnetische,heusler1903,sciadv.1602241}. Conversely, when the atomic number of $X$ becomes lower than $Y$ ($\Z_X<\Z_Y$), the structure adapts it by swapping their atomic positions, and an inverse Heusler is formed~\cite{PhysRevB.98.094410,PhysRevB.97.060406}. Even when one of the sublattices is vacant, the structure still remains stable and the material is called half-Heusler ($XYZ$)~\cite{7b02685,PhysRevB.82.085108}. Thus, Heusler alloys, endowed with structural flexibility, are fostering the development of a new class of materials. Hence, very recently, a new subclass of Heusler alloys was engineered by substituting $X$ with the $Z$ element and vice versa in the full-Heusler structure. It becomes a special kind of alloy, where $p$-block elements increase to 50\% of the constituents, and the formula unit is reversed, $Z_2XY$ (from $X_2YZ$). Hence, this new subclass of Heusler alloys is termed hereafter as `anti-Heusler' alloys.

Since a large number of transition metals and $p$-block elements in the periodic table leverage a plethora of forming a diverse array of compounds within the playground of versatile tunable atomic arrangements, a desired magnetic and electronic property can be achieved in Heusler alloys by conducive band structure~\cite{200601815}. The magnetic properties of these alloys are one of the pivotal interests, as understanding the complex interplay between magnetism and electronic structure is beneficial for spintronics applications. The magnetic moment of Heusler alloys is perceived with the help of the well-known S-P rule~\cite{PhysRev.49.931,PhysRev.54.899}, which delineated that the magnetic moment, $m_t$, will be the difference between the total valence electron count (VEC) and 24 (18), for four-atomic full-, quaternary-, inverse-(three atomic half-) Heuslers ~\cite{10.1063/1.2167629,GRAF20111}. Although the rule is well established for all known subclasses of Heusler compounds, it has not been explored yet for newly discovered anti-Heusler alloys, where the enhanced (50\%) $p$-block element may make a substantial difference. According to the standard S-P rule, the (four-atomic) anti-Heusler compound, having VEC 24, is not anticipated to exhibit any magnetic moment. Till now, no VEC 24 anti-Heusler has been reported and all the reported compounds are known to manifest ferromagnetic ordering, whereas the ordered spin structure gets destroyed at low temperatures, paving the way to reentrant spin/cluster glass behaviour~\cite{PhysRevB.78.134406,PhysRevB.97.184421,Samanta2022}. In this work, the formation and magnetic properties of a new anti-Heusler alloy, Al$_2$MnCu, with VEC 24, are reported. The structural characterization encompasses x-ray and neutron diffraction, along with extended X-ray absorption fine structure reveals that the compound forms in B2-type  with 5\% Cu-Al disorder structure. The magnetic measurements complemented with the low-temperature neutron diffraction study elucidate the compound undergoes a magnetic ordering, more importantly, manifests a large magnetic moment of $\sim$1.8 $\mu_B$/f.u., despite carrying VEC 24. A phenomenological model based on molecular orbital hybridization is also proposed, in line with the formalism used earlier for understanding the S-P rule, to explain the observed deviation from such standard rule in Al$_2$MnCu.

\section{Experimental details}
\begin{figure*}
\centering
\includegraphics[width=1.0\textwidth]{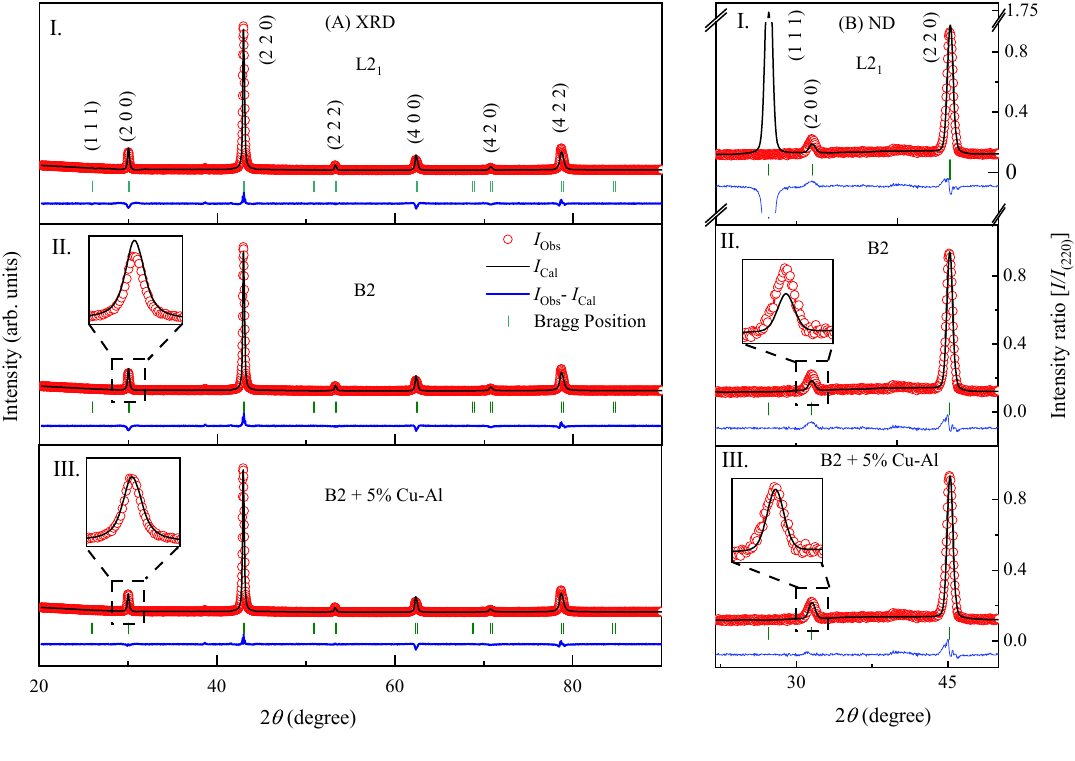}
\caption{Rietveld refinement of the (A) X-ray and (B) neutron diffraction pattern assuming L2$_1$-ordered (I), B2-type disordered (II), and B2 + 5\% Cu-Al disordered structure (III).}
\label{fig:XRD ND_comparison}
\end{figure*}

The polycrystalline sample of Al$_2$MnCu was synthesized by the arc-melting technique. Highly pure ($>$99.9 wt.~\%, Alfa Aesar, USA) constituent elements with a stoichiometric ratio were melted in an inert gas atmosphere on a water-cooled copper hearth. The weight loss due to evaporation of elemental Manganese was compensated by taking an extra 2\% weight of Mn. The phase purity of the synthesized material was confirmed by the wavelength dispersive spectroscopy-based electron probe micro-analysis (EPMA) technique (Model: EPMA -1720 HT,  SHIMADZU, Japan). The crystal structure of the sample was investigated through powder X-ray diffraction (XRD), using a commercial diffractometer (rotating Cu anode, 9 kW, Model: TTRAX-III, Rigaku Corp., Japan) at various temperatures between 15-300~K. The crystal and magnetic structure is probed using neutron diffraction (ND) measurements at the ECHIDNA beam-line ($\lambda$ = 1.622 \AA ) in ANSTO, Australia~\cite{Avdeev:ks5608}, at temperatures of 3~K and 500~K. The crystal and magnetic structures were determined by performing Rietveld refinement of the XRD and ND spectra using the FULLPROF software package~\cite{RODRIGUEZCARVAJAL199355}. X-ray Absorption Spectroscopy (XAS) measurements, which comprises both X-ray Near Edge Structure (XANES) and Extended X-ray Absorption Fine Structure (EXAFS) techniques, have been carried out to probe the local structure of Al$_2$MnCu at Mn and Cu $K$-edge. The energy-scanning EXAFS beamline (BL-9) at the Indus-2 Synchrotron Source (2.5 GeV, 100 mA) at Raja Ramanna Centre for Advanced Technology (RRCAT), Indore, India~\cite{10.1063/1.4872706,Basu_2014}was used to perform the measurement. The XAS measurements have been performed in the transmission mode for standards and in fluorescence mode for Al$_2$MnCu samples. The detailed information regarding the beamline and measurement setup for both transmission and fluorescence modes is described elsewhere~\cite{doi:10.1021/acsaem.3c02617}.

Magnetic measurements were carried out utilizing a commercial superconducting quantum interference device (SQUID) magnetometer in the temperature range of 2$–$400~K and magnetic field range of 0-70~kOe (Quantum Design, Inc., USA). Temperature-dependent magnetization measurements were done by employing zero-field-cooled (ZFC) and field-cooled (FC) protocols. In the ZFC method, the sample was initially cooled from the highest temperature without applying any magnetic field whereas in FC protocol, the same was done in the presence of a specified applied magnetic field. For both protocols,  the magnetization data were collected during heating from 2~K  to 400~K in the presence of the same magnetic field, under which the system temperature was brought down during the FC process~\cite{PhysRevB.106.224427}. The isothermal magnetization with a variation of the magnetic field was performed at different temperatures. At each measurement, the sample was cooled down to the specified temperature from the paramagnetic temperature region in the absence of any external magnetic field.

\section{Results and Discussions}
\subsection{Structural details}
\label{Structural details}
\subsubsection{X-ray diffraction}
The X-ray diffraction (XRD) pattern of Al$_2$MnCu is shown in Fig.~\ref{fig:XRD ND_comparison} A. The diffraction pattern can be well indexed by the L2$_1$ ordered structure with lattice parameter $a = b = c =$ 5.954(1) \AA \, and space group $Fm\bar{3}$m (No. 225), where Al, Mn and Cu atoms occupy at 8$c$ (0.25, 0.25, 0.25), 4$b$ (0.5, 0.5, 0.5) and 4$a$ (0, 0, 0) Wyckoff sites, respectively. However, due to the nearby constituent elements, Heusler alloys often deviate from the perfect L2$_1$ ordered structure and exhibit atomic disorder in their respective Wyckoff positions~\cite{PhysRevB.69.144413,10.1063/1.1513216}. The atomic disorder, restricted in between 4$a$ and 4$b$ Wyckoff sites, is a partly disorder and called as B2-type, whereas full disorder occurs when all atoms are randomly distributed in all three Wyckoff positions, and this higher disorder is known as A2-type disorder~\cite{PhysRevB.108.054430,PhysRevB.108.245151}. These disorders are reflected in the intensity of (111) and (200) Bragg peaks with respect to (220). The disappearance of the (111) peak intensity suggests a B2-type disorder in the system. Despite the absence of (111) peak in the XRD pattern of Al$_2$MnCu, the spectra can be well described by the ordered L2$_1$ structure (Fig.~\ref{fig:XRD ND_comparison} A I). This discrepancy can be explained by considering the X-ray scattering factors of the constituent atoms, $viz.$, Cu and Mn and the structure factor equations of the characterizing Bragg peaks (111), (200), and (220), which can be expressed as~\cite{PhysRevB.99.104429,CHAKRABORTY2024173215}
\begin{eqnarray}\label{eq:sfa2mc}
F_{(111)} &=& 4|f_{Cu}-f_{Mn}|\nonumber\\
F_{(200)} &=& 4|f_{Cu}+f_{Mn}-2f_{Al}| \\
F_{(220)} &=& 4|f_{Cu}+f_{Mn}+2f_{Al}|\nonumber.
\end{eqnarray}
The above equation indicates that the intensity of the (111) peak would depend on the differences in the scattering factors of Cu and Mn atoms and can result in vanishing intensity even for perfectly ordered structure due to the similar X-ray scattering factor of these atoms. As the absence of (111) peak in the diffraction pattern for Heusler alloys suggests a B2-type disordered structure, accordingly, we have also attempted to fit the XRD spectra with this disordered structure (Fig.~\ref{fig:XRD ND_comparison} A II). Most of the fit parameters as well as goodness of fits remain nearly unchanged for both the fits. Thus, the XRD data fails to determine the internal atomic arrangements within the crystal structure due to close values of X-ray scattering form-factors of Mn and Cu, although it confirms the single-phase nature of the newly synthesized anti-Heusler compound Al$_2$MnCu.
\subsubsection{Neutron diffraction}
\begin{figure}[h]
\centering
\includegraphics[width=01\columnwidth]{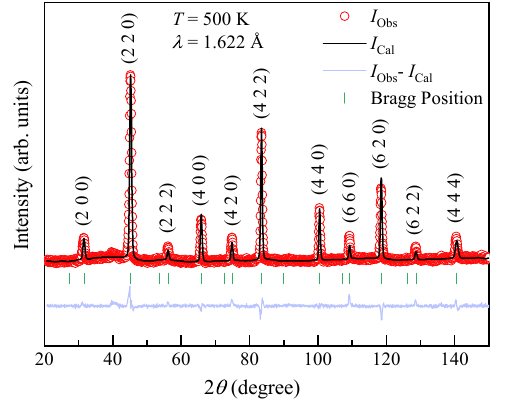}
\caption{Rietveld refinement of the neutron diffraction pattern taken at 500 K of Al$_2$MnCu with B2 and 5\% Cu-Al disorder.}
\label{fig:ND 500K}
\end{figure}

\begin{figure}
  \centering
  \includegraphics[width=0.7\columnwidth]{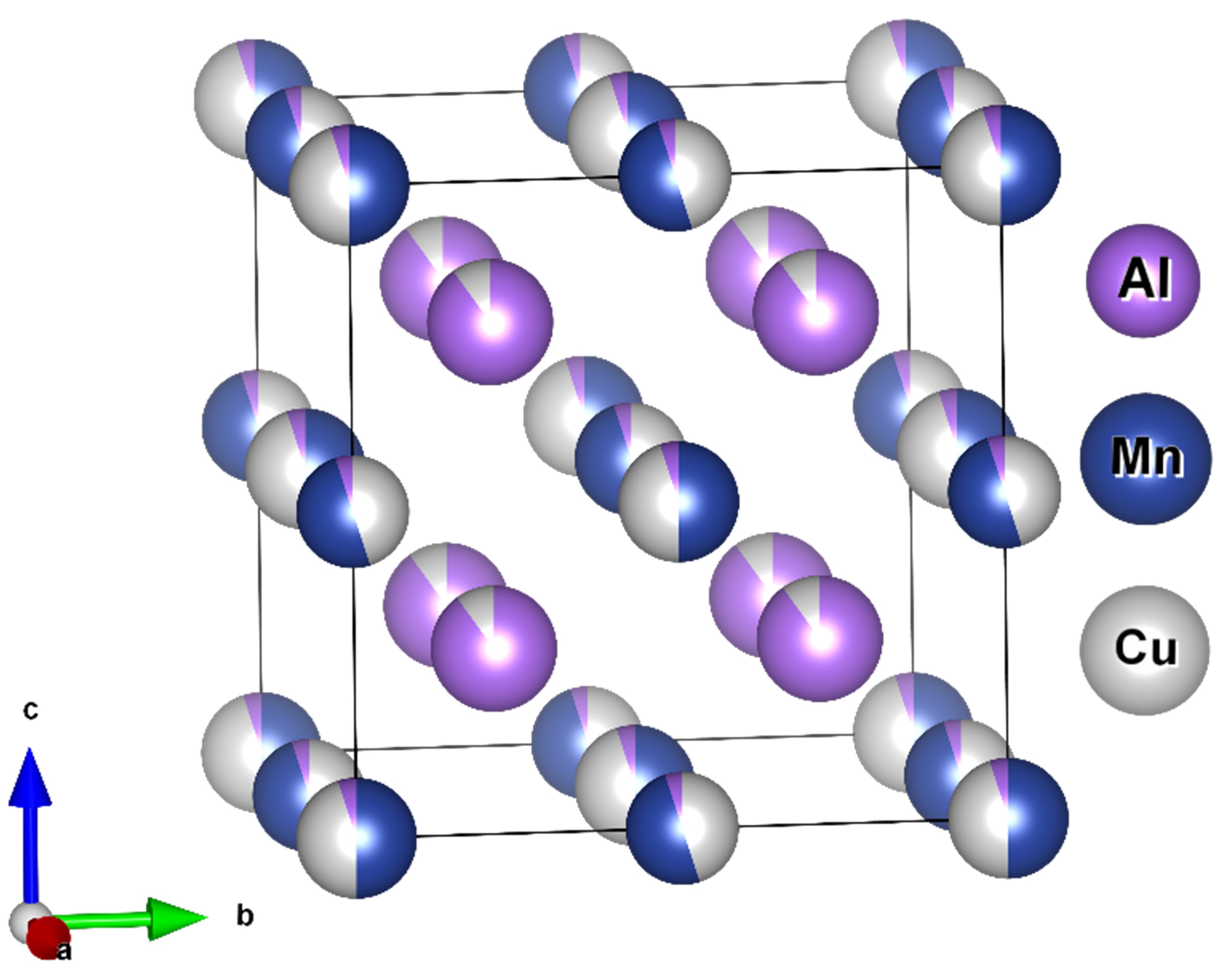}
  \caption{Crystal structure of Al$_2$MnCu having B2-type with extra 5\% Cu-Al disorder.}\label{fig:Crystal structure}
  \label{fig: Al2MnCu_crystal structure}
\end{figure}

In contrast to X-ray scattering factors that increase monotonously with atomic numbers, the neutron scattering length of elements (as well as their isotopes) are not proportional to their respective atomic numbers, where neighbouring elements in the periodic table may exhibit vastly different scattering lengths. This property thus can be of great utility in resolving such structural ambiguity provided by XRD analysis. As the neutron coherent scattering lengths of Cu ($\sim~$7.718 fm) and Mn ($\sim$ -3.73 fm) are of opposite sign and differs widely, the (111) Bragg peak intensity in the neutron diffraction (ND) data would be of considerable value in case of an ordered structure. Accordingly, the ND measurements were done for the compound and the spectra measured at $T = 500$~K are shown in Fig.~\ref{fig:ND 500K} along with full-Rietveld analysis. As evident from the figure, in our attempt to fit the ND data with L2$_1$ ordered structure, the theoretically generated peak intensity of (111) peak appears to be almost double that of (220) Bragg peak (Fig.~\ref{fig:XRD ND_comparison}, B I), as the neutron coherent scattering length is positive for Cu and Al, while it is negative for Mn. However, the non-existence of such a peak in the experimental ND pattern signifies the presence of B2-type disorder in the system~\cite{PhysRevB.84.224416}. Consequently, the ND spectra were analysed using a B2-type disordered structure as shown in Fig.~\ref{fig:XRD ND_comparison} B II. The goodness of fit enhanced greatly in the B2-structural model. Although a weak mismatch still exists in (200) peak intensity, which can not be explained by considering A2-type disorder, as this would reduce the fitted intensity further, yielding a larger difference between the observed and fitted spectra. Additional analysis suggests that the larger peak intensity is the result of a small (5\%) additional atomic disorder between Al from tetrahedral and Cu from octahedral sites (Fig.~\ref{fig:XRD ND_comparison} B III and Fig.~\ref{fig:ND 500K}). The overall ND pattern at $T = 500$~K can thus be well indexed with lattice parameters $a = b = c =$ 5.968(2)\AA\ considering B2-type disorder along with 5\% Cu-Al disorder. The XRD pattern can also be well explained considering these atomic arrangements, as shown in Fig.~\ref{fig:XRD ND_comparison} A III. The detailed atomic occupancies are presented in Table~\ref{tab:Occupancy}.

\begin{table}[h]
\begin{tabular}{lcccccc}
\hline
                      & \multicolumn{1}{l}{} & \multicolumn{1}{l}{} & \multicolumn{1}{l}{}                                 & \multicolumn{1}{l}{} & \multicolumn{1}{l}{} & \multicolumn{1}{l}{}                                        \\
Site                  &                      &                      & Element                                              &                      &                      & Occupancy                                                   \\
                      & \multicolumn{1}{l}{} & \multicolumn{1}{l}{} & \multicolumn{1}{l}{}                                 & \multicolumn{1}{l}{} & \multicolumn{1}{l}{} & \multicolumn{1}{l}{}                                        \\ \hline
                      & \multicolumn{1}{l}{} & \multicolumn{1}{l}{} & \multicolumn{1}{l}{}                                 & \multicolumn{1}{l}{} & \multicolumn{1}{l}{} & \multicolumn{1}{l}{}                                        \\
4$a$ (0,0,0)          &                      &                      & \begin{tabular}[c]{@{}c@{}}Mn\\ Cu\\ Al\end{tabular} &                      &                      & \begin{tabular}[c]{@{}c@{}}0.500\\ 0.453\\ 0.047\end{tabular} \\
                      &                      &                      &                                                      &                      &                      &                                                             \\
4$b$ (0.5,0.5,0.5)    &                      &                      & \begin{tabular}[c]{@{}c@{}}Mn\\ Cu\\ Al\end{tabular} &                      &                      & \begin{tabular}[c]{@{}c@{}}0.500\\ 0.453\\ 0.047\end{tabular} \\
                      &                      &                      &                                                      &                      &                      &                                                             \\
8$c$ (0.25,0.25,0.25) &                      &                      & \begin{tabular}[c]{@{}c@{}}Al\\ Cu\end{tabular}      &                      &                      & \begin{tabular}[c]{@{}c@{}}0.953\\ 0.047\end{tabular}       \\
                      & \multicolumn{1}{l}{} & \multicolumn{1}{l}{} & \multicolumn{1}{l}{}                                 & \multicolumn{1}{l}{} & \multicolumn{1}{l}{} & \multicolumn{1}{l}{}                                        \\ \hline\hline
\end{tabular}
\caption{Site occupancy of Al$_2$MnCu obtained by analysing the neutron diffraction pattern.}
\label{tab:Occupancy}
\end{table}

\subsubsection{Extended x-ray absorption fine structure}
\begin{figure}
  \centering
  \includegraphics[width=01\columnwidth]{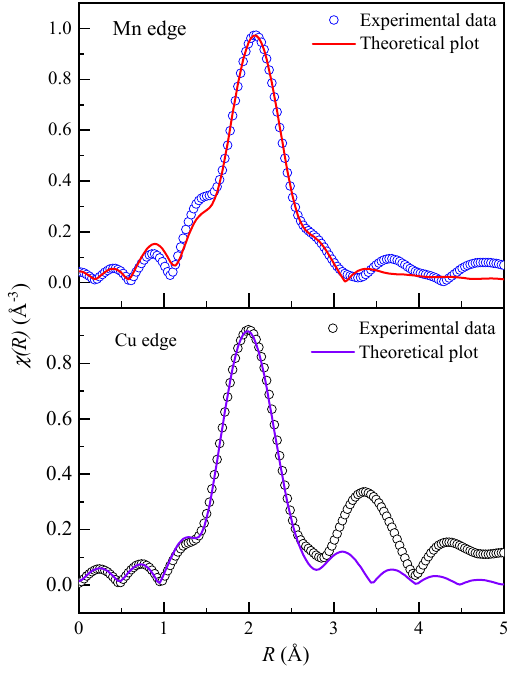}
  \caption{Fourier transformed EXAFS spectra of Al$_2$MnCu taken
at the Mn edge (upper), and the Cu edge (lower).}\label{fig:exaf}
\end{figure}

Unlike XRD and ND, which provide overall structural information, the extended X-ray absorption fine structure (EXAFS) measurement probes the local environment of a particular constituent atom of a compound. To obtain the qualitative information about the local structure around the Cu and Mn sites in the samples from the EXAFS data, the experimentally obtained $\mu(E)$ versus  $E$ spectra of the respective edges have been reduced and Fourier transformed to obtain the radial distribution functions or $\chi(R)$ versus $R$  plots and these are subsequently fitted with theoretically generated radial distribution functions. The ATHENA and ARTEMIS codes available within the Demeter software package have been used in the above analysis~\cite{NEWVILLE1995154}. Fig.~\ref{fig:exaf} shows the experimentally obtained $\chi(R)$ versus $R$ data along with the best fit theoretical plots. The EXAFS fitting is done simultaneously at both the edges in the range of 1.0$-$3.2 \AA\ in $R$ space using the co-ordination number ($N$), distance ($R$) between the respective atomic pairs and the disorder factor ($\sigma^2$), known as Debye–Waller factor as the fitting parameters, the best-fit parameters being presented in Table~\ref{tab:Exaf}. The scattering paths used to create the theoretical plot have been obtained from the crystal structure of Al$_2$MnCu obtained from XRD and ND measurements. The goodness of fit in each case has been determined by the value of the $R_{factor}$~\cite{Kelly2008}. The rather low value of the goodness of fit parameter ($R_{factor} \sim$ 0.007) indicates a structural description close to reality.
\begin{table*}[]
\begin{tabular}{clclclclclclclc}
\hline \hline
\multicolumn{7}{c}{\multirow{2}{*}{Mn edge}}                                                                                                                      &                      & \multicolumn{7}{c}{\multirow{2}{*}{Cu edge}}                                                                                                                      \\
\multicolumn{7}{c}{}                                                                                                                                              &                      & \multicolumn{7}{c}{}                                                                                                                                              \\ \hline
\multirow{2}{*}{Path} & \multicolumn{1}{c}{} & \multirow{2}{*}{R(\AA)} & \multicolumn{1}{c}{} & \multirow{2}{*}{N}   & \multicolumn{1}{c}{} & \multirow{2}{*}{$\sigma^2$} & \multicolumn{1}{c}{} & \multirow{2}{*}{Path} & \multicolumn{1}{c}{} & \multirow{2}{*}{R(\AA)} & \multicolumn{1}{c}{} & \multirow{2}{*}{N}   & \multicolumn{1}{c}{} & \multirow{2}{*}{$\sigma^2$} \\
                      &                      &                      &                      &                      &                      &                        &                      &                       &                      &                      &                      &                      &                      &                        \\ \cline{1-7} \cline{9-15}
\multicolumn{7}{l}{}                                                                                                                                              &                      & \multicolumn{7}{l}{}                                                                                                                                              \\
Mn-Al                 &                      & 2.54$\pm$0.01             &                      & 8.0$\pm$0.24             &                      & 0.0065$\pm$0.001           &                      & Cu-Al                 &                      & 2.53$\pm$0.01            &                      & 8.0$\pm$0.24             &                      & 0.027$\pm$0.001            \\
\multicolumn{1}{l}{}  &                      & \multicolumn{1}{l}{} &                      & \multicolumn{1}{l}{} &                      & \multicolumn{1}{l}{}   &                      & \multicolumn{1}{l}{}  &                      & \multicolumn{1}{l}{} &                      & \multicolumn{1}{l}{} &                      & \multicolumn{1}{l}{}   \\
Mn-Cu                 &                      & 2.85$\pm$0.01            &                      & 3.0$\pm$0.09             &                      & 0.001$\pm$0.0005           &                      & Cu-Cu                 &                      & 2.64$\pm$0.01            &                      & 3.0$\pm$0.09             &                      & 0.001$\pm$0.0005           \\
\multicolumn{1}{l}{}  &                      & \multicolumn{1}{l}{} &                      & \multicolumn{1}{l}{} &                      & \multicolumn{1}{l}{}   &                      & \multicolumn{1}{l}{}  &                      & \multicolumn{1}{l}{} &                      & \multicolumn{1}{l}{} &                      & \multicolumn{1}{l}{}   \\
Mn-Mn                 &                      & 3.05$\pm$0.01            &                      & 3.0$\pm$0.09             &                      & 0.0015$\pm$0.0008          &                      & Cu-Mn                 &                      & 2.85$\pm$0.01            &                      & 3.0$\pm$0.09             &                      & 0.0011$\pm$0.0005          \\
\multicolumn{1}{l}{}  &                      & \multicolumn{1}{l}{} &                      & \multicolumn{1}{l}{} &                      & \multicolumn{1}{l}{}   &                      & \multicolumn{1}{l}{}  &                      & \multicolumn{1}{l}{} &                      & \multicolumn{1}{l}{} &                      & \multicolumn{1}{l}{}   \\ \hline \hline
\end{tabular}
\caption{Bond length $R$, coordination number $N$, and Debye-Waller or disorder factor $\sigma^2$ obtained by EXAFS fitting for Al$_2$MnCu at the Mn and Cu edges.}
\label{tab:Exaf}
\end{table*}

For a well-ordered L2$_1$ structure, the first coordination shell of both Mn and Cu would be filled by eight Al atoms, whereas the second coordination shell of Mn (Cu) would be inhabited by six Cu (Mn) atoms. In the case of B2 disorder, as the octahedral sites (4$a$ and 4$b$) are occupied by both Mn and Cu atoms with 50$-$50 occupancy, unlike L2$_1$ for the Mn (Cu) edge, the second coordination shell is not filled with six Cu (Mn) atoms, rather it will be occupied with a 50$-$50 ratio, $i.e.$, three Mn and three Cu. The analysis of the Fourier transformed spectra of both the Mn and Cu edges reveals that the first coordination shell is filled by eight Al atoms, whereas the six sites of the second coordination shell are occupied by three Mn and three Cu atoms, as expected for the B2 disorder structure. Thus, the EXAFS result also corroborated quite well with the B2-type disorder in the crystal structure as obtained from the analysis of XRD and ND spectra.
\subsubsection{Electron probe micro analysis (EPMA)}
\begin{figure}[ht]
\centering
\includegraphics[width=01\columnwidth]{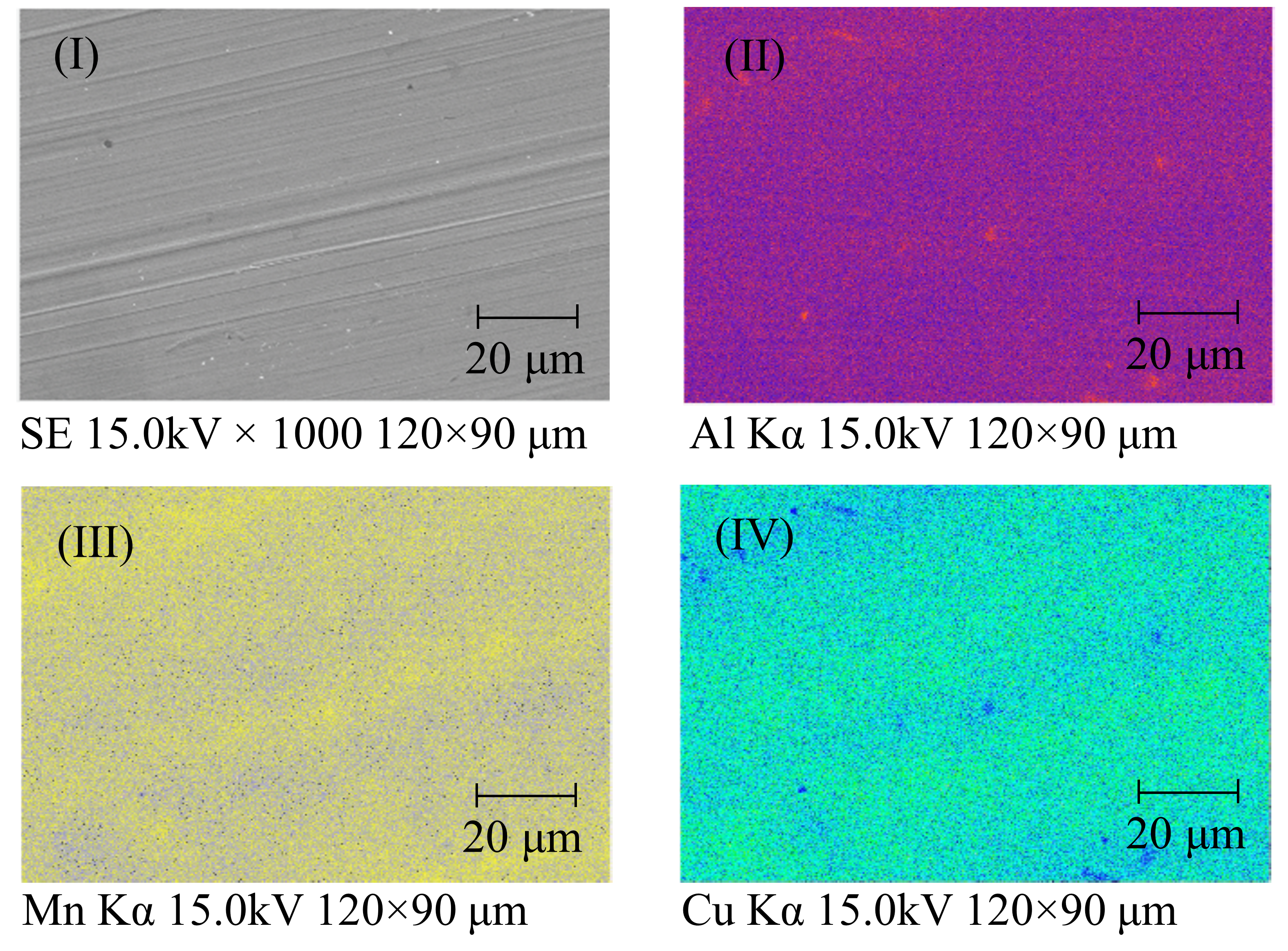}
\caption{EPMA results with homogeneous distribution of (I) Al$_2$MnCu, and elemental mapping of (II) Al, (III) Mn and (IV) Cu.}
\label{fig:EPMA}
\end{figure}
%\vspace{10pt}
\begin{table}[ht]
\begin{tabular}{|c|c|c|}
\hline
Element & Weight \% & Atomic \% \\ \hline
Al      & 29.2     & 47.5     \\ \hline
Mn      & 33.2      & 26.5     \\ \hline
Cu      & 37.4     & 25.9     \\ \hline
\end{tabular}
\caption{Atomic composition of the compound determined from EPMA measurement.}
\label{tab:composition}
\end{table}
The phase purity of the sample has also been checked by electron probe microanalysis (EPMA). The uniform intensity of the image mapped in the back-scattered electron (BSE) spectroscopy mode (Fig.~\ref{fig:EPMA} I) indicates the homogeneous formation of the compound. The single-phase nature is further confirmed by the K-edge mapping of individual constituent elements on the sample surface (Figs.~\ref{fig:EPMA} II, III, and IV). The atomic percentage obtained from the analysis is given in Table~\ref{tab:composition} within an experimental error of $\sim3\%$.

\subsubsection{Low-temperature XRD}
\begin{figure}
\includegraphics[width=01\columnwidth]{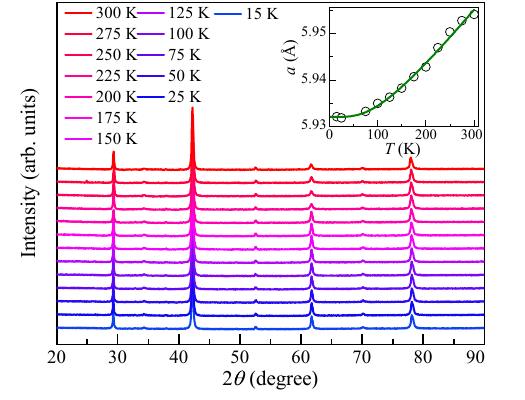}
\caption{Low-temperature XRD pattern of Al$_2$MnCu. Inset shows temperature dependence lattice parameter. Solid line is the fitting with Debye model.}
\label{fig:LTXRD}
\end{figure}

Some members of different Heusler alloys family are known to exhibit structural transition at low temperatures~\cite{PhysRevB.97.184421,10.1063/1.4757951}. Such information is quite essential for any magnetic Heusler alloys to find the magnetic spin structure by analysing low-temperature ND spectra. Accordingly, the XRD measurements of Al$_2$MnCu were extended in the temperature region 300-15~K and the diffraction patterns are shown in Fig.~\ref{fig:LTXRD}. All the spectra conform to $Fm\bar{3}m$ space group, suggesting the absence of any structural transition within this temperature range for the studied compound. The temperature dependence of lattice parameter thus estimated follows usual lattice contraction (Fig.~\ref{fig:LTXRD} (inset)) and can be fitted with the standard Bloch-Gr\"{u}neisen model~\cite{PhysRevB.94.104414}:  $a(T)= \gamma U(T)/K_0 + a_0$, where $a_0$ is the lattice parameter at $T$ = 0 K, $K_0$ is known as the Young's modulus and $\gamma$ is the Gr\"{u}neisen parameter. $U(T)$ is the internal energy which can be determined from the Debye approximation, $U(T)= 9NK_BT\left(\frac{T}{\theta_D}\right)^3 \int_{0}^{\frac{\theta_D}{T}}\frac{x^3}{e^x-1}\,dx$. Using this equation Debye temperature has been estimated as $\theta_D$ = 398 K. Similar values of Debye temperatures have earlier been reported in many Heusler alloys~\cite{PhysRevB.72.184434}.

\subsection {Magnetic properties}
\label{sec:magnetization}
\subsubsection{Magnetic ordering}
\begin{figure}
\centering
\includegraphics[width=01\columnwidth]{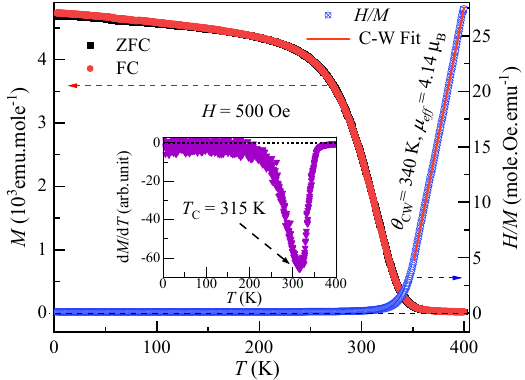}
\caption{ Magnetization vs. Temperature  plot (ZFC and FC) at 500 Oe applied field (left scale) and $H/M$ vs temperature curve (right scale). The temperature dependent first order derivative of FC Magnetization curve w.r.t. $T$ is plotted in the inset.}
\label{fig:MT}
\end{figure}
To understand the nature of the magnetic ground state of Al$_2$MnCu, the temperature dependence of magnetization measurements has been carried out under different applied magnetic fields in the temperature range 2-400~K, under ZFC and FC conditions. Fig.~\ref{fig:MT} shows the magnetic behaviour measured at the applied field, $H$ = 500~Oe. The compound exhibits a long-range ferromagnetic ordering below $T_{\rm C} = 315$~K, determined from the dip in d$M(T)$/d$T$ behaviour~\cite{PhysRevB.105.214407}. The paramagnetic region ($T >$ 355 K) of the magnetization curve can be well described by the Curie-Weiss (CW) law, $\chi = C/(T-\theta_{CW}$), where $C = N_ A\mu_{eff}^2\mu_B^2/3K_B$, is Curie constant and $\theta_{CW}$ is the paramagnetic Curie-Weiss temperature, as depicted in the Fig.~\ref{fig:MT}~\cite{PhysRevB.107.094421} . The best fit of the $H/M(T)$ curve using CW analysis yielded $\mu_{eff}$ to be  4.14~$\mu_B/f.u.$ and $\theta_{CW}$ $\sim$340~K, signifying the presence of dominant ferromagnetic interaction in the system between the Mn spins. In most of the standard ferromagnetic systems, $\theta_{CW}$ is generally found to be slightly larger than $T_{\rm C}$, which is also the case here. Since the members of the VEC 24 family of Heusler alloys are not expected to carry any magnetic moment, let alone any magnetic ordering, whereas a large value of $\mu_{eff}$ and the ferromagnetic ordering of Al$_2$MnCu appear to counter the trend.
\begin{figure}[h]
\centering
\includegraphics[width=01\columnwidth]{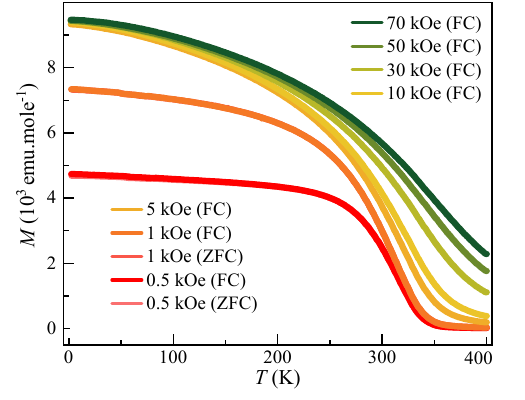}
\caption{Thermo-magnetic curve at different applied field. The curve gets broadened in the transition region with an increasing applied filed.}
\label{fig:MT_all}
\end{figure}

The long-range nature of ferromagnetic ordering is further confirmed when the magnetization measurements have been extended with larger values of externally applied magnetic fields, as shown in Fig.~\ref{fig:MT_all}. The $M(T)$ around the Curie temperature of Al$_2$MnCu get broadened along with an enhancement in $T_{\rm C}$ with increasing $H$, as observed in typical long-range ferromagnetic system.
\subsubsection{Isothermal Magnetization and violation of S-P rule}
\label{Sec:Isothermal magnetization}

\begin{figure}
\centering
\includegraphics[width=01\columnwidth]{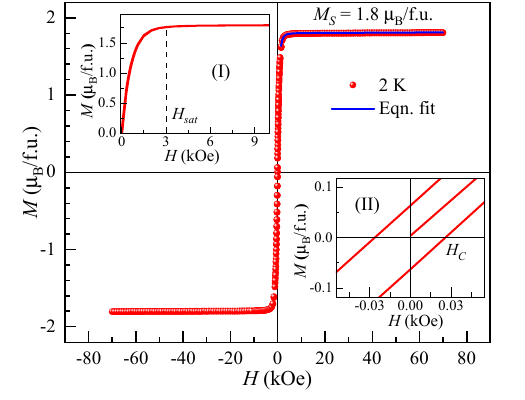}
\caption{Isothermal magnetization curve measured at 2~K. The curve at high field region is fitted with an approach to saturation law. Inset (I) shows the saturating field, $H_{Sat}$ to be, $\sim$3~kOe and (II) indicates low coercivity, $H_C\sim$250~Oe, of the material.}
\label{fig:MH2K}
\end{figure}

\begin{figure}
\centering
\includegraphics[width=01\columnwidth]{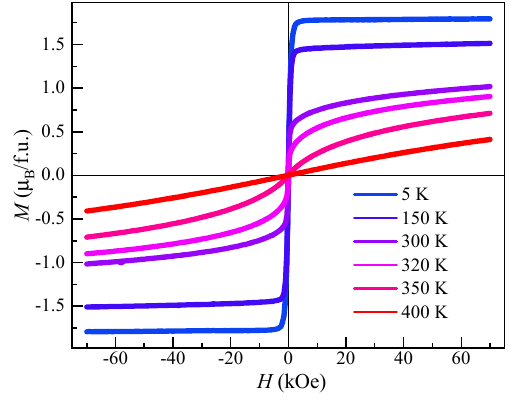}
\caption{Isothermal Magnetization curve  measured at different temperatures. Linear curve, measured at 400~K, suggests paramagnetic region.}
\label{fig:MH_all}
\end{figure}

The temperature dependence of magnetization however gets saturated for a field of $<$ 5~kOe, suggesting the system to have rather low anisotropy. Magnetic field dependence of isothermal magnetization measurement is shown in Fig.~\ref{fig:MH2K} for the lowest measured temperature, $T = 2$~K and in Fig.~\ref{fig:MH_all} for different other temperatures of interest. The small coercivity ($H_C$ $\sim$25~Oe) suggests the soft ferromagnetic nature of the system. The magnetisation however gets saturated at a lower field value of $H_{Sat}\sim$3~kOe, suggesting the system to have rather low magnetocrystalline anisotropy associated with the FM domains. The $M(H)$ data shows a non-linear behaviour up to 350~K, suggesting the development of short-range correlations extended above its Curie temperature. This is also in conformation with the non-linear behaviour of $H/M(T)$ curve below 350~K. The saturation magnetization ($M_S$) at $T$ = 2~K is determined by fitting the $M(H)$ curve using an approach to saturation law~\cite{principles}, $M= M_S \left(1-\frac{A}{H}- \frac{B}{H^2}\right)+\chi H$, where $M_S$ is the saturation magnetization, the parameters $A$ and $B$ are used to explain the effect of structural defects and magneto-crystalline anisotropy, respectively, whereas $\chi$ is attributed to the high-field susceptibility. The estimated value of $M_S$ is 1.8$~\mu_{\rm B}$/f.u. In a long-range ordered ferromagnetic system, the ordered moment $M \sim M_S$, thus is this case $\sim 1.8~\mu_{\rm B}$/f.u. Noteworthy, according to the S-P rule, generally applied for the Heusler alloys, the ordered moment for the studied compound Al$_2$MnCu is expected to be zero as the total VEC is 24. However, the observation of a robust ferromagnetic ground state with a large ordered moment suggests that the standard Slater-Pauling mechanism based on the conventional orbital hybridization mechanism is not exactly compatible with this anti-Heusler alloy Al$_2$MnCu. A phenomenological model is provided in Sec.~\ref{S-P rule} to explain the deviation.

\subsubsection{Low temperature neutron diffraction}
\label{sec:MagND}

\begin{figure}
\centering
\includegraphics[width=01\columnwidth]{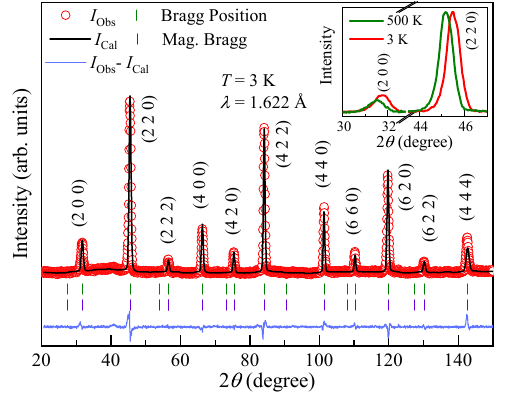}
\caption{Rietveld refinement of the neutron diffraction pattern taken at 3~K. The inset shows the extended region around (200) and (220) Bragg peak positions for both the neutron diffraction spectra taken at 3~K and 500~K. A small increase in (200) Bragg peak intensity in the 3~K spectra indicates the magnetic ordering in the system.}
\label{fig:ND_3K}
\end{figure}
\begin{figure}
\includegraphics[width=0.3\textwidth]{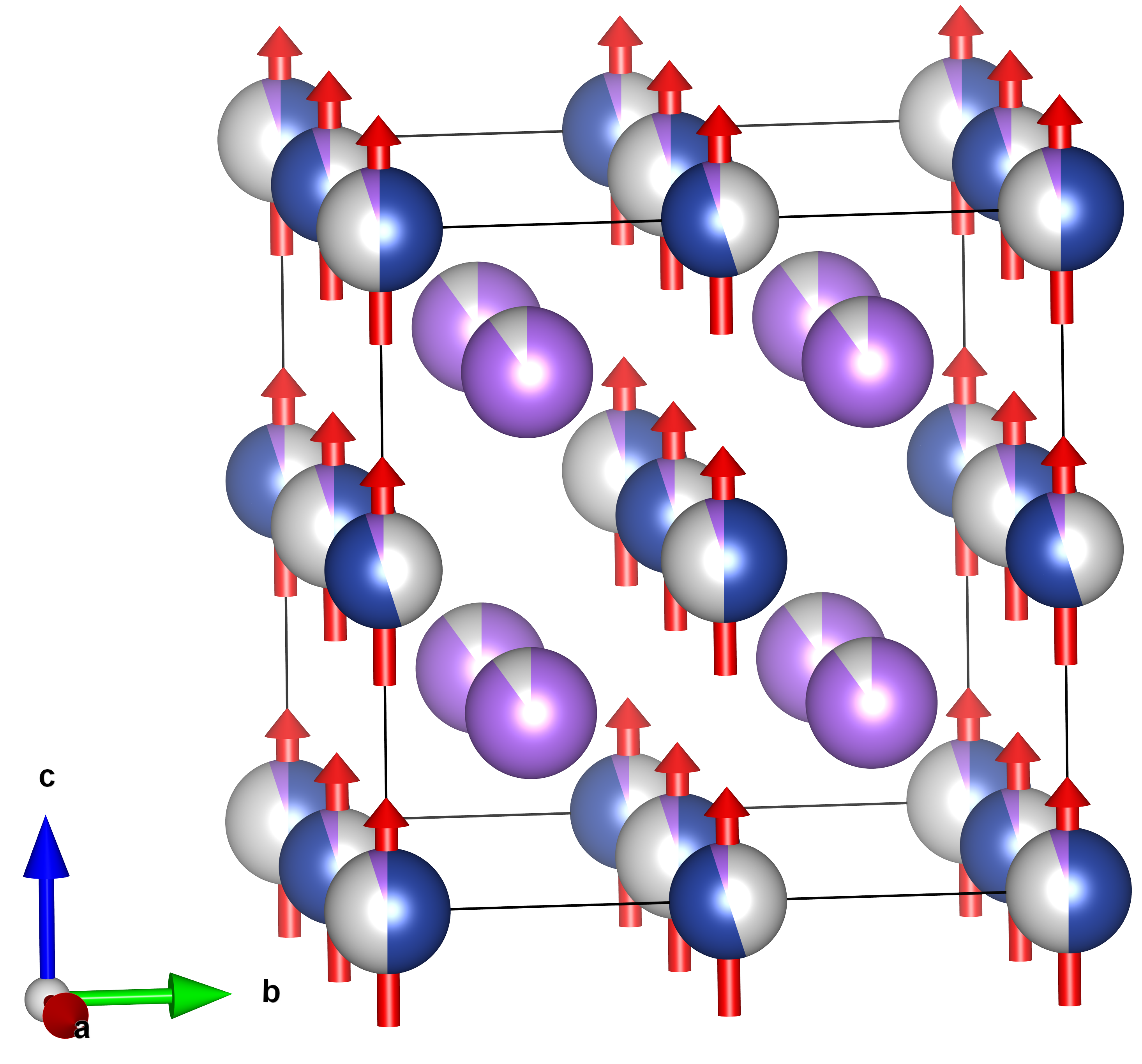}
  \caption{Magnetic spin structure of Al$_2$MnCu, determined from ND analysis at $T$= 3~K. The colour scheme of the constituent atoms have already been presented in Fig.~\ref{fig: Al2MnCu_crystal structure}.The spins are associated with Mn atoms.}
  \label{Fig:magnetic structure}
\end{figure}
Neutron diffraction (ND) measurements further confirm the long-range ferromagnetic order observed in Al$_2$MnCu. The ND spectra measured at $T$ = 3~K, $i.e.$, below the Curie temperature of the compound, are shown in Fig.~\ref{fig:ND_3K}. In comparison with the ND spectra taken at $T$ = 500 K (Fig.~\ref{fig:ND 500K}), no additional Bragg peak appears, thus ruling out the possibility of any anti-ferromagnetic ordering, which is also consistent with the magnetometry data. A close examination of ND spectra, however, reveals an enhancement of (200) lattice Bragg peak intensity at 3~K, confirming a long-range ferromagnetic ordering (propagation vector, $k$ = (0, 0, 0)) in the system. Rietveld analysis of ND spectra a 3~K suggests that Mn is the only moment-carrying atom which is equally distributed to 4$a$ and 4$b$ sites with a value of 0.859(2)~$\mu_B$ at each of these two sites (Fig.~\ref{Fig:magnetic structure}). The total moment thus turns out to be 1.718~$\mu_B$/Mn, closely matching with the saturation moment value of $\sim~$1.8$~\mu_B$/f.u. obtained from the isothermal magnetisation measurements (Sec.~\ref{Sec:Isothermal magnetization}). The result, in turn, firmly confirms that Al$_2$MnCu, contradicting the standard S-P rule, possesses a high magnetic moment instead of zero.

\subsection{A phenomenological model to explain the deviation of Slater-Pauling rule in Al$_2$MnCu}
\label{S-P rule}

%In many Heusler alloy systems, the presence of atomic disorders is often considered to be the primary reason for the violation of the S-P rule. Although most such cases are involved with a reduction in magnetic moment~\cite{PRXEnergy.3.033006,PhysRevB.94.214423,PhysRevB.69.144413,KOGACHI2009723}, disorder-induced enhancement of moment have also been observed in some materials~\cite{PhysRevB.62.3296,PhysRevB.98.205130}. For example in a VEC 24 full-Heusler alloy, it was found that to achieve a similar increase in the magnetic moment, the system not only needs to have complete swap-disorder (A2-type) among constituent atoms~\cite{Garmroudi2022}, it is also to be complemented by a small amount of additional elemental substitution~\cite{PhysRevB.107.014108}. If the swap-disorder is only partial (B2-type),  and the system remains unperturbed by any additional elemental substitution, the magnetic moment could be raised to at most by $\sim$0.22 $\mu_B$/f.u. only~\cite{Garmroudi2022,PhysRevB.62.3296,PhysRevB.80.235121,PhysRevB.98.205130,DATTA2021167522}. As the VEC 24 anti-Heusler alloy Al$_2$MnCu forms with B2-type disorder only, it is quite unlikely that the same disorder alone can yield a magnetic moment as high as $\sim$1.8 $\mu_B/f.u$.

The violation of the S-P rule for Al$_2$MnCu raises a much broader question: whether it is restricted to this particular compound only or generally expected for all members of anti-Heusler alloys. Accordingly, we revisited the origin of the S-P rule applicable to Heusler alloys. Slater and Pauling proposed a relation between the total magnetic moment, m$_t$, and the valence electron count for 3$d$ elements and their binary alloys~\cite{PhysRev.49.931,PhysRev.54.899}. Subsequently, the relation was extended to Heusler alloys where it was described that the total magnetic moment of the system can be considered as a summation of the individual contribution of each constituent element, $\sum_{n}(N_V-6)$, where N$_V$ is the valence electron count of the respective elements.

\begin{figure}
\centering
\includegraphics[width=01\columnwidth]{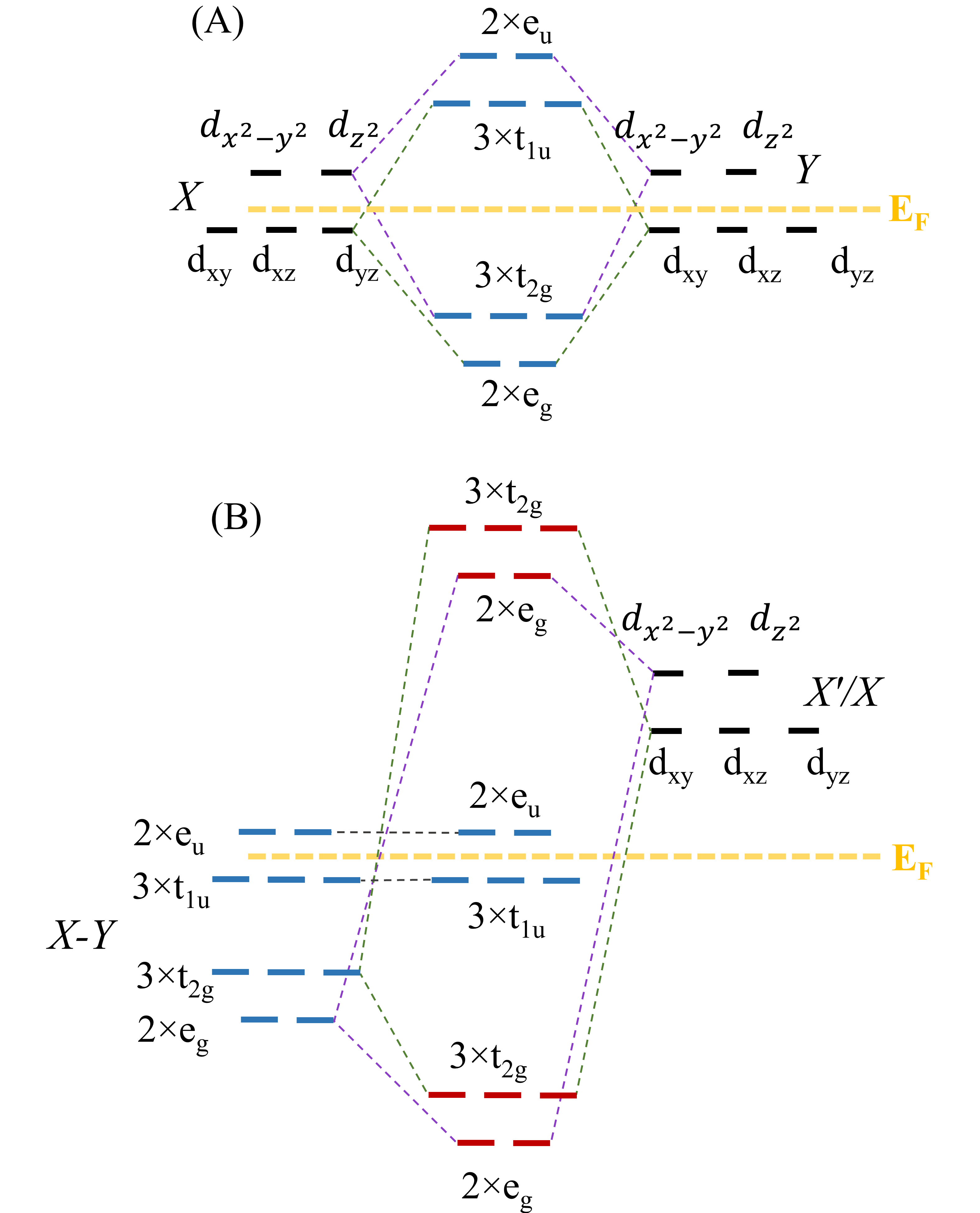}
\caption{Possible hybridization between minor spin orbital of transition metals of (A) half and anti-Heusler, (B) full- Heusler alloys.}
\label{fig:Hybrdization of full and half}
\end{figure}

The validity of the rule has already been established through the molecular orbital approach, for both the three-atomic ($XYZ$) as well as four-atomic ($X_2YZ/XX'YZ$) Heusler alloys~\cite{PhysRevB.66.174429,PhysRevB.87.024420,PhysRevB.90.064408,Galanakis_2006,GRAF20111,Belkhir2023}. In case of $XYZ$ type of compounds (half-Heusler), hybridization takes place between the 3$d$ orbitals of $X$ and $Y$ atoms, whereas $s$ and $p$-orbitals, of $Z$ atoms, lie far below the Fermi energy. The double-degenerated $d$ orbitals ($d_{x^2-y^2}$ and $d_{z^2}$) of both $X$ and $Y$ hybridize and create two high energy anti-bonding orbitals, $e_u$, along with two low energy bonding orbitals $e_g$  (Fig.~\ref{fig:Hybrdization of full and half} (A)). On the other hand, hybridization between the triple-degenerated $d$ orbitals ($d_{xy}, d_{xz}$ and  $d_{yz}$) of $X$ and $Y$ yield three anti-bonding ($3\times t_{1u}$) and three bonding ($3\times t_{2g}$) orbitals  and Fermi energy ($E_{\rm F}$) is located between them. The $sp$-orbitals (1 $\times s$ and 3 $\times p$) of $p$-block element ($Z$), lie much below $E_{\rm F}$, and for clarity they are not presented in figure. Thus, a total of nine states, below the $E_{\rm F}$, can be accommodated by nine minor spins and nine major spins as well. If the total valance electron count (VEC or, $N_V$) is more than 18, the excess electrons occupy the anti-bonding ($t_u$ and $e_u$) orbitals and are devoid of any accompanying opposite spins. These unbalanced major spins are responsible for determining the magnetic moment value of the compound. On the contrary, for VEC $<$ 18, although all the nine states below $E_{\rm F}$ are accompanied by minor spins, there are not enough major spins to counter-balance, and the magnetism is contributed by those unpaired minor spins. The difference between VEC and 18, be it positive or negative, thus determines the magnetic moment in $XYZ$ type of Heusler alloys, and the S-P rule becomes, $m_t = |N_V-18|\,\mu _B /f.u.$

In contrast to half-Heusler ($XYZ$), the four-atomic, $i.e.$, full- ($X_2YZ$), inverse- ($Y_2XZ$) or quaternary- ($XX^{\prime}YZ$) Heusler alloys, contain one more transition metal element. The involvement of the additional $d$-orbitals belonging to this third transition element ($X/Y/X^{\prime}$) results in modification of the hybridization process by introducing a further step, $i.e.$, the hybridization now becomes a two-fold process. While the first step of hybridization between $X$ and $Y$ atoms remain essentially similar to that described above (Fig.~\ref{fig:Hybrdization of full and half} (A)), the second step involves further hybridization of $d$ states of $X^{\prime}$ atoms with the bonding states of $X$-$Y$. The final outcome is two sets (bonding and anti-bonding) of double-degenerated $e$ and triple-degenerated $t$ states (Fig.~\ref{fig:Hybrdization of full and half} (B)).

Here, only the bonding states ($t_{2g}$ \& $e_{g}$) forming after the first hybridization step take part in the second step. The anti-bonding states ($e_u$ \& $t_{1u}$) remain passive in this process but spread across the Fermi level. The complete hybridization process thus yields a total of eight $d$ orbitals ($3\times t_{1u}$, $3\times t_{2g}$ and $2\times e_{g}$), in addition to four $sp$ orbitals of $Z$-element (not taking part in any of these hybridization steps), below $E_{\rm F}$. Thus, a total of 12 states below $E_{\rm F}$, having occupancy of 12 major (and 12 minor), yielding, $m_t = |N_V-24|\,\mu_B/f.u.$

Although anti-Heusler alloys ($Z_2XY$) form with four atoms per formula unit, owing to the presence of excess $p$ block element in place of transition metal, the above description of regular four-atomic Heuslers cannot be applied. As the $p$ block elements lie far below the $E_{\rm F}$ and hence do not take part in the hybridization process, the hybridization scheme between the remaining two transition elements will be exactly identical to half-Heuslers (Fig.~\ref{fig:Hybrdization of full and half} (A)) with additional four $sp$ orbital of the extra $Z$ element will be located far below $E_{\rm F}$. In other words, the molecular orbital approach to anti-Heusler suggests the development of five bonding (below $E_{\rm F}$) and five anti-bonding (above $E_{\rm F}$) states, whereas the two $Z$ atoms contributes to eight $sp$-orbitals, also much below $E_{\rm F}$ (Fig.~\ref{fig:Hybrdization of full and half} (A)). This scheme thus dictates a total of 13 (= 5+8) states below $E_{\rm F}$. Since each of these states can accommodate 1 minor as well as 1 major spin, a total of 26 electrons would be required to fully compensate the magnetism in anti-Heusler alloys, in contrast to other four-atomic Heusler alloys, $viz.$, full-, inverse- and quaternary-, where valence electron count is required to be 24. Therefore, in Al$_2$MnCu-types of anti-Heusler alloys with the majority of $p$-electron elements, the magnetic moment should be best described by the relation, $m_t = |N_V-26|\,\mu_B/f.u$. Since the valence electron count of this compound is 24, the proposed orbital hybridization model suggests a maximum ordered magnetic moment of 2 $\mu_B/f.u.$ for Al$_2$MnCu. The experimentally observed magnetic moment value of $m_t\sim~$1.8$~\mu_B/f.u.$ is quite well in agreement with this value. The small deviation from the theoretical value of 2~$\mu_B/f.u.$ could be attributed to structural disorder present in the system~\cite{PRXEnergy.3.033006,PhysRevB.94.214423,PhysRevB.69.144413,KOGACHI2009723}. However, despite the apparent agreement with the experimental data, the model still remains as conjectural and speculative only, till it can be tested for many more anti-Heusler compounds having a wide variation of their respective VEC numbers. One also should not discard other possible sources, for example, atomic disorder-induced magnetic moments, as had been reported in Fe$_2$VAl~\cite{Garmroudi2022,PhysRevB.107.014108}, where the degree of disorder is controlled by quenching the system from different higher temperatures. Since the scope of such investigation can not be encompassed within the present work, the exact origin of the magnetic moment in VEC 24 anti-Heusler alloy Al$_2$MnCu can not be identified with absolute certainty.

\section{Conclusion}
In this study, we present the synthesis, structural and magnetic properties of a novel anti-Heusler alloy, Al$_2$MnCu. It forms in B2 with 5\% Cu-Al disorder, as revealed through XRD, ND and EXAFS measurements. It is the only known member of the anti-Heusler family with a VEC 24, a configuration for which the conventional S-P rule predicts a non-magnetic ground state. In contrast, our temperature and magnetic field-dependent magnetization measurements combined with neutron diffraction studies reveal that the compound exhibits a long-range ferromagnetic ordering above room temperature ($T_{\rm C}\sim$ 315 K), that too with substantial saturation magnetic moment of $\sim~$1.8$~\mu_{\rm B}$/f.u. In order to explain the origin of high magnetic moment in Al$_2$MnCu having VEC 24 and plausible deviation from S-P rule, a phenomenological model based on molecular hybridization mechanism is proposed for $Z_2XY$-type anti-Heusler compounds. The proposed model closely aligns with the experimental observations in Al$_2$MnCu and is likely to explain the magnetic moment of related anti-Heusler alloys.

\begin{acknowledgments}
Soumya Bhowmik would like to sincerely acknowledge SINP, India for the fellowship and research facility. The author acknowledge the Central Research Facility (CRF) of Indian Institute of Technology Delhi for EPMA measurements and Ashis Kundu for fruitful discussions. R.C. acknowledge Ames National Laboratory, which is supported and operated for the U.S. Department of Energy by Iowa State University under Contract No. DE-AC02-07CH11358.
\end{acknowledgments}

\normalem
\bibliographystyle{apsrev4-2}

\end{document}